\documentclass{svmult}

\usepackage{makeidx}     
\usepackage{graphicx}    
\usepackage{multicol}    
\usepackage{url}
\usepackage{amsmath,amssymb,epsfig, }
\usepackage{latexsym}
\newcommand{\spec}              { \ensuremath {\mbox{{\rm spec}}} }
\renewcommand{\span}              { \ensuremath {\mbox{{\rm span}}} }
\newcommand{\cond}              { \ensuremath {\mbox{{\rm cond}}} }

\def\sign#1{\hbox{\rm \,sign}(#1)}

\newcommand{\eq}[1]{eq.~(\ref{#1})}


\newcommand{\eqnref}[1]{(\ref{#1})}

\begin{document}

\title*{Numerical Methods for the QCD Overlap Operator:
       II. Optimal Krylov Subspace Methods}
\titlerunning{Optimal Krylov Methods}
\author{Guido Arnold\inst{1}\and Nigel Cundy\inst{1}\and
Jasper van den Eshof\inst{2}\and
Andreas Frommer\inst{3}\and Stefan Krieg\inst{1}\and Thomas Lippert\inst{1} \and Katrin Sch\"afer\inst{3} }
\authorrunning{G. Arnold et al.}
\makeatletter
\institute{Department of Physics,
      University of Wuppertal, Germany
\texttt{\{arnold,cundy,krieg,lippert\}@theorie.physik.uni-wuppertal.de}
\and Department of Mathematics, University of D\"usseldorf,
Germany \texttt{eshof@am.uni-duesseldorf.de}
\and Department of Mathematics, University of Wuppertal,
Germany \texttt{\{frommer,schaefer\}@math.uni-wuppertal.de}
}
\makeatother
%
%
\maketitle

\begin{abstract}
We investigate optimal choices for the (outer) iteration
method to use when solving linear systems with Neuberger's overlap
operator in QCD. Different formulations for this operator give rise
to different iterative solvers, which are optimal for the respective
formulation. We compare these methods in theory and practice
to find the overall optimal one.
For the first time, we apply the so-called
SUMR method of Jagels and Reichel to the shifted unitary version
of Neuberger's operator, and show that this method is in a sense
the optimal choice for propagator computations. When solving
the ``squared'' equations in a dynamical simulation with two degenerate flavours, it turns out that
the CG method should be used.
\end{abstract}

\section{Introduction}
\label{sec:1}

Recently, lattice formulations of QCD  respecting chiral symmetry
have attracted a lot of attention. A particular promising such
formulation, the so-called overlap fermions, has been proposed in
 \cite{Narayanan:2000qx}. From
the computational point of view,  we have
to solve linear systems involving the sign function $\sign{Q}$ of the (hermitian)
Wilson fermion matrix $Q$. These computations are very costly, and it
is of vital importance to devise efficient numerical schemes.

A direct computation of $\sign{Q}$ is not feasible, since $Q$ is large and
sparse, whereas $\sign{Q}$ would be full. Therefore,
numerical algorithms have to follow  an
inner-outer paradigm: One performs an outer Krylov subspace
method where each iteration requires the computation of a matrix-vector
product involving $\sign{Q}$. Each such product is computed through
another, inner iteration using matrix-vector multiplications with $Q$.
In an earlier paper \cite{EFL02} we investigated methods for the inner iteration and
established the Zolotarev rational approximation together with the
multishift CG method \cite{Glassner:1996gz} as the method of choice.

In the present paper we investigate optimal methods for the outer iteration. We consider
two situations: the case of a propagator computation and the case of a pseudofermion
computation within a dynamical hybrid Monte Carlo simulation where one has to solve the ``squared'' system.
As we will see, the optimal method for the case of a propagator computation is a
not so well-known method due to Jagels and Reichel \cite{JR94}, whereas in the case of the
squared system it will be best to apply classical CG on that squared system rather than
using a two-pass approach.

This paper is organized as follows: We first introduce our notation in Section~\ref{sec:not}.
We then discuss different equivalent formulations of the Neuberger overlap
operator and establish useful relations between the eigenpairs of
these different formulations  (Section~\ref{sec:sys}).
We discuss optimal Krylov subspace methods for the various systems
in Section~\ref{sec:met} and give some theoretical results on their convergence
speed based on the spectral information from Section~\ref{sec:sys}.
In Section~\ref{sec:comp} we compare the convergence speeds both,
theoretically and in practical experiments.
Our conclusions will be summarized in Section~\ref{sec:con}.

\section{Notation} \label{sec:not}
The Wilson-Dirac fermion operator,
\begin{equation}
M = I -\kappa D_W,
\end{equation}
represents a nearest neighbour coupling on a four-dimensional
space-time lattice, where the ``hopping term'' $D_W$ is a
non-normal sparse matrix, see \eqnref{eq:Diracmatrix} in the appendix. The
coupling parameter $\kappa$ is a real number which is related to the
bare quark mass.

The massless overlap operator is defined as
\[
   D_0 = I + M \cdot (M^{\dagger}M)^{-\frac{1}{2}}.
\]
For the massive overlap operator, for notational convenience we use
a mass parameter $\rho > 1$ such that this operator is given as
\begin{equation} \label{overlap:def}
   D = \rho I + M \cdot (M^{\dagger}M)^{-\frac{1}{2}}.
\end{equation}
In the appendix, we explain that this form is just a scaled
version of Neuberger's original choice, and we relate $\rho$ to the
quark mass, see \eqnref{MASSDEFINITION} and \eqnref{FERMIONMASS}.

Replacing $M$ in \eqnref{overlap:def} by its hermitian form $Q$,
see \eqnref{HWD}, the overlap operator can equivalently be written as
\begin{equation} \label{overlap2:def}
D = \rho I + \gamma_5 \sign{Q} = \gamma_5 \cdot (\rho \gamma_5 + \sign{Q}),
\end{equation}
with $\gamma_5$ being defined in \eqnref{GAMMA5} and $\sign{Q}$ being the
standard matrix sign function. Note that $\rho \gamma_5 + \sign{Q}$ is
hermitian and indefinite, whereas $\gamma_5 \sign{Q}$ is unitary.

To reflect these facts in our notation, we define:
$$
D_u = \rho I + \gamma_5 \sign{Q},\qquad
D_h = \rho \gamma_5 + \sign{Q},
$$
where $D_u = \gamma_5 D_h$. Both these
operators are normal, i.e.\ they commute with their adjoints.

\section{Formulations and their spectral properties} \label{sec:sys}
\subsection{Propagator computations}
When computing quark propagators, the systems to solve are of the form
\begin{equation} \label{unitary_eq}
D_u x = ( \rho I + \gamma_5\sign{Q}) x = b.
\end{equation}

Multiplying this shifted unitary form by $\gamma_5$,
we obtain its {\em hermitian
indefinite form}  as
\begin{equation} \label{hermitian_eq}
D_h x = (\rho \gamma_5 + \sign{Q})x = \gamma_5 b.
\end{equation}

The two operators $D_u$ and $D_h$
are intimately related. They are both normal, and as a consequence the eigenvalues (and the
weight with which
the corresponding eigenvectors appear in the initial residual) solely govern the
convergence behaviour of an optimal Krylov subspace method used to solve
the respective equation. 

Very interestingly,
the eigenvalues of the different operators can be explicitly related
to each other. This allows a quite detailed discussion of the convergence properties
of adequate Krylov subspace solvers.
To see this, let us introduce an auxiliary decomposition for $\sign{Q}$: Using the chiral representation,
the matrix $\gamma_5$ on the whole lattice can be represented as a $2 \times 2$
block diagonal matrix
\begin{equation}\label{gamma5blockeq}
\gamma_5 = \left( \begin{array}{cc} I & 0 \\
                                    0 & -I 
                 \end{array}
            \right) ,
\end{equation}
where both diagonal blocks $I$ and $-I$ are of the same size. Partitioning $\sign{Q}$
correspondingly gives
\begin{equation}\label{signblockeq}
\sign{Q} = \left( \begin{array}{cc} S_{11} & S_{12} \\
                                    S^{\dagger}_{12} & S_{22}
                 \end{array}
            \right).
\end{equation}
In Lemma~\ref{dec_lem} below we give a convenient decomposition for this matrix that is closely related to
the so-called {\em CS decomposition} (see \cite[Theorem~2.6.3]{GLo96} and the references
therein), an important tool in matrix analysis.
Actually, the lemma may be regarded as a variant of the CS decomposition for 
hermitian matrices where the decomposition here can be achieved using a
similarity transform. 
The proof follows the same lines as the proof for the existence of the CS decomposition given in \cite{PWe94}.
\begin{lemma}\label{dec_lem}
There exists a unitary matrix $X$ such that
\[
\sign{Q} = X
 \left( \begin{array}{cc} \Phi & \Sigma \\
                                    \Sigma & \Psi
                 \end{array}
            \right) X^{\dagger},
	    \quad 
	    \hbox{ with }
X =  \left( \begin{array}{cc} X_1 & 0 \\
                                    0 & X_2
                 \end{array}
            \right).	    
\]
The matrices $\Phi$, $\Psi$ and $\Sigma$ are real and diagonal with diagonal elements
$\phi_j$, $\psi_j$ and $\sigma_j\ge 0$, respectively.
Furthermore, $\phi_j^2 +\sigma_j^2 = \psi_j^2 + \sigma_j^2 = 1$ and
\begin{eqnarray}
& \phi_j = -\psi_j \quad &\hbox{if} \quad \sigma_j > 0 \label{sigmagt_eq}\\
& \phi_j, \psi_j \in \{-1,+1\}   \quad &\hbox{if}\quad \sigma_j = 0. \label{sigmaeq_eq}
\end{eqnarray}
\end{lemma}

Note that in the case of \eqnref{sigmagt_eq} we know that $\phi_j$ and $\psi_j$
have opposite signs, whereas in case \eqnref{sigmaeq_eq} we might have
$\phi_j = \psi_j = 1$ or $\phi_j = \psi_j = -1$.
The key point for $X$ in Lemma~\ref{dec_lem}
is the fact that $\gamma_5 X = X \gamma_5$. This allows us to relate the
eigenvalues and -vectors of the different
formulations for the overlap operator  via $\phi_j$, $\psi_j$ and
$\sigma_j$.  In this manner we give results complementary to
 Edwards et al. \cite{EHN98}, where relations between the eigenvalues
(and partly the -vectors)
of the different formulations for the overlap operator were given without
connecting them to $\sign{Q}$ via $\phi_j$, $\psi_j$ and
$\sigma_j$.
The following lemma gives expressions for the eigenvectors and -values of the
shifted unitary operator $D_u$.
\begin{lemma} \label{specdu_lem}
With the notation from Lemma \ref{dec_lem}, let
      $x_j^{1}$ and $x_j^{2}$ be the $j$-th column of $X_1$ and $X_2$,
      respectively.
Then $\spec(D_u) = \{ \lambda_{j,\pm}^u \}$ with
      \begin{eqnarray*}
          & \lambda_{j,\pm}^u = \rho + \phi_j \pm i\sqrt{1-\phi^2_j}
                 &\quad\mbox{ if }\quad \sigma_j \not = 0 \\
         & \lambda_{j,+}^u = \rho + \phi_j, \; \lambda_{j,-}^u = \rho - \psi_j
        &\quad\mbox{ if }\quad \sigma_j = 0.
       \end{eqnarray*}
        The corresponding eigenvectors are
  \begin{equation}
    \begin{array}{ll}
  z^u_{j,\pm} =  \left( \begin{array}{c}
                         \mp i x_j^{1} \\
                         x_j^{2}
                         \end{array}
                  \right)
    & \quad\mbox{ if }\quad \sigma_j \not = 0\\
     z^u_{j,+} = \left( \begin{array}{c}
                          x_j^{1} \\
                          0
                         \end{array}
                  \right), \; 
   z^u_{j,-} = \left( \begin{array}{c}
                          0 \\
                          x_j^2
                         \end{array}
                  \right) &\quad \mbox{ if }\quad \sigma_j = 0.
    \end{array}
\end{equation}
\end{lemma}
\begin{proof}
With the decomposition for $\sign{Q}$ in
Lemma~\ref{dec_lem}, we find,
using  $X^{\dagger} \gamma_5  =  \gamma_5 X^{\dagger}$,
\[
X ^{\dagger}\gamma_5\sign{Q} X =  \gamma_5 X^{\dagger} \sign{Q} X \\
    =   \left( \begin{array}{cc} \Phi & \Sigma \\
  -\Sigma & -\Psi
  \end{array}
  \right). 
\]
Since the actions of $X$ and $X^{\dagger}$ represent a similarity transform,
we can easily
derive the eigenvalues and eigenvectors of $\gamma_5\sign{Q}$ from the matrix
on the right.
This can be accomplished by noticing that this matrix can be permuted
to have a block diagonal form with $2\times 2$ blocks on the diagonal, so that the
problem reduces to the straightforward computation of
 the eigenvalues and eigenvectors of $2 \times 2 $ matrices. This concludes the proof.
\end{proof}

As a side remark, we mention that Edwards et al. \cite{EHN98} observed that the eigenvalues of $D_u$ can be efficiently
computed by exploiting the fact that most of the eigenvalues come in
complex conjugate pairs. Indeed, using Lemma~\ref{dec_lem} and the fact that
$\gamma_5 X = X \gamma_5$  we see that we only
have to compute the eigenvalues of the hermitian matrix $S_{11}$ (and to check for
one or minus one eigenvalues in $S_{22}$).

With the same technique as in Lemma~\ref{specdu_lem} we can also 
find expressions for the eigenvalues and eigenvectors of the hermitian indefinite
formulation.
\begin{lemma} \label{specdh_lem}
With the same notation as in Lemma~\ref{specdu_lem}, we have that $\spec(D_h) = \{\lambda^{h}_{j,\pm}\}$ with
\begin{eqnarray*}
  & \lambda^{h}_{j,\pm} = \pm \sqrt{1 + 2 \phi_j \rho + \rho^2} &\quad \mbox{ if }\quad
          \sigma_j \not = 0 \\
 & \lambda^{h}_{j,+} = \rho + \phi_j, \; \lambda^{h}_{j,-} =
       -\rho + \psi_j &\quad \mbox{ if } \quad\sigma_j = 0.
\end{eqnarray*}
The corresponding eigenvectors are
\begin{eqnarray*}
  z^{h}_{j,\pm} =  \left( \begin{array}{c}
  (\phi_j  \rho + \lambda_{j,\pm}^{h})x_j^{1} \\
  \sqrt{1-\phi_j^2} \, x_j^{2}
                 \end{array}
            \right) \; &\quad\mbox{ if }\quad \sigma_j \not = 0,\\
 z^{h}_{j,+} =  \left( \begin{array}{c}
                    x_j^{1} \\
                    0
                 \end{array}
            \right), \;
z^{h}_{j,-} =  \left( \begin{array}{c}
                       0  \\
              x_j^{2}
                 \end{array}
            \right)\; &\quad\mbox{ if }\quad \sigma_j = 0.
\end{eqnarray*}
\end{lemma}

\begin{figure}
\centering
\includegraphics[scale=0.33]{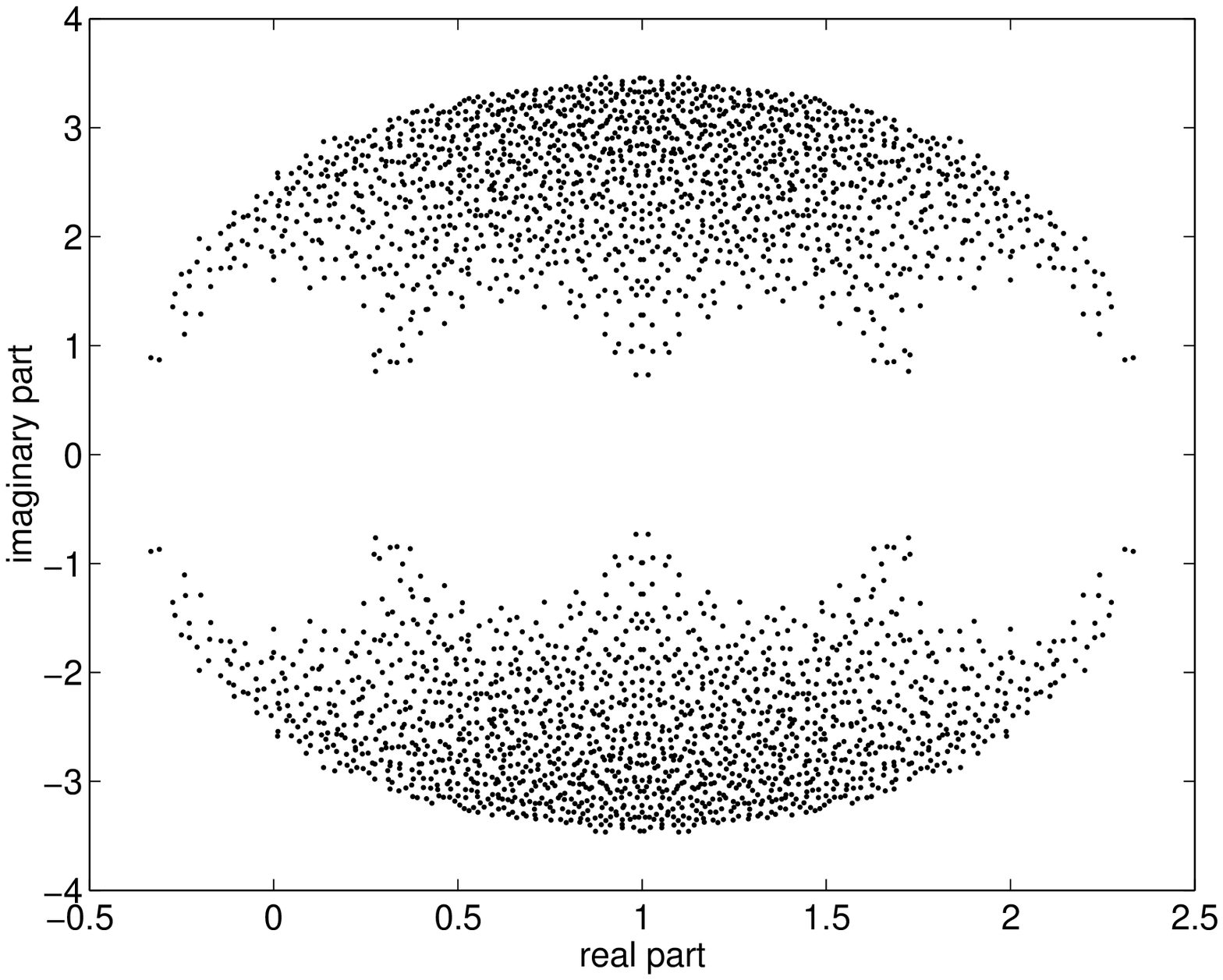}
\includegraphics[scale=0.33]{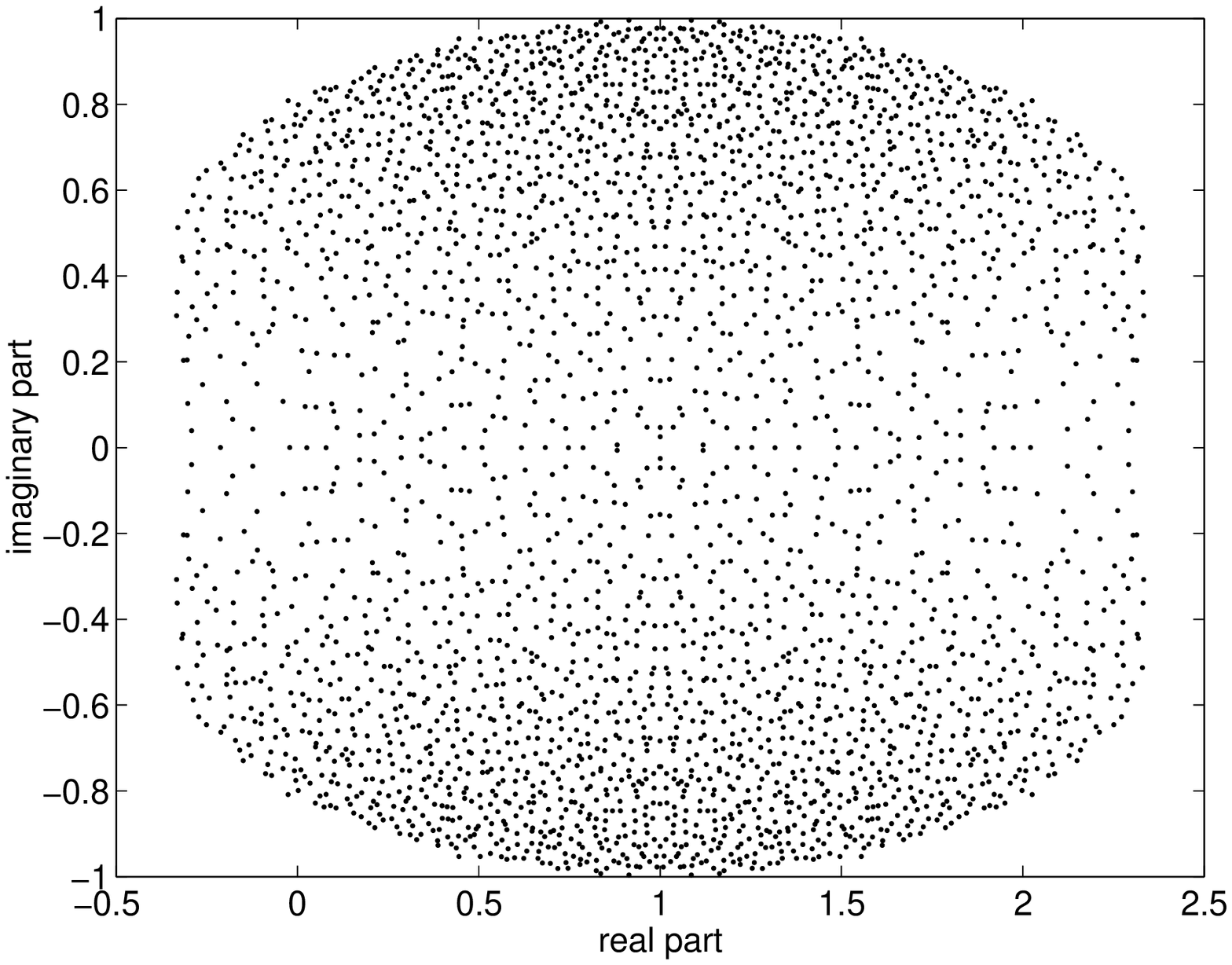}
\caption{Eigenvalues for the Wilson fermion matrix $M$ for configurations A (left) and B (right)
}
\label{Wilson_fig}
\end{figure}

As an illustration to the results of this section, we performed numerical
computations for two sample configurations. Both are on a $4^4$ lattice with $\beta = 6.0$ (configuration A) and
$\beta = 5.0$ (configuration B) respectively.
\footnote{The hopping matrices for both configurations are available at www.math.uni-wuppertal.de/org/SciComp/preprints.html as well as \texttt{matlab} code for all methods presented here.
The configurations are also available at Matrix Market as {\tt conf6.0\_00l4x4.2000.mtx} and {\tt conf5.0\_00l4x4.2600.mtx} respectively.}
Figure~\ref{Wilson_fig} shows plots of the eigenvalues of $M$
for these configurations. We used $\kappa = 0.2129$ for configuration A, $\kappa = 0.2809$ for configuration B.

\begin{figure}
\centering
\includegraphics[scale=0.33]{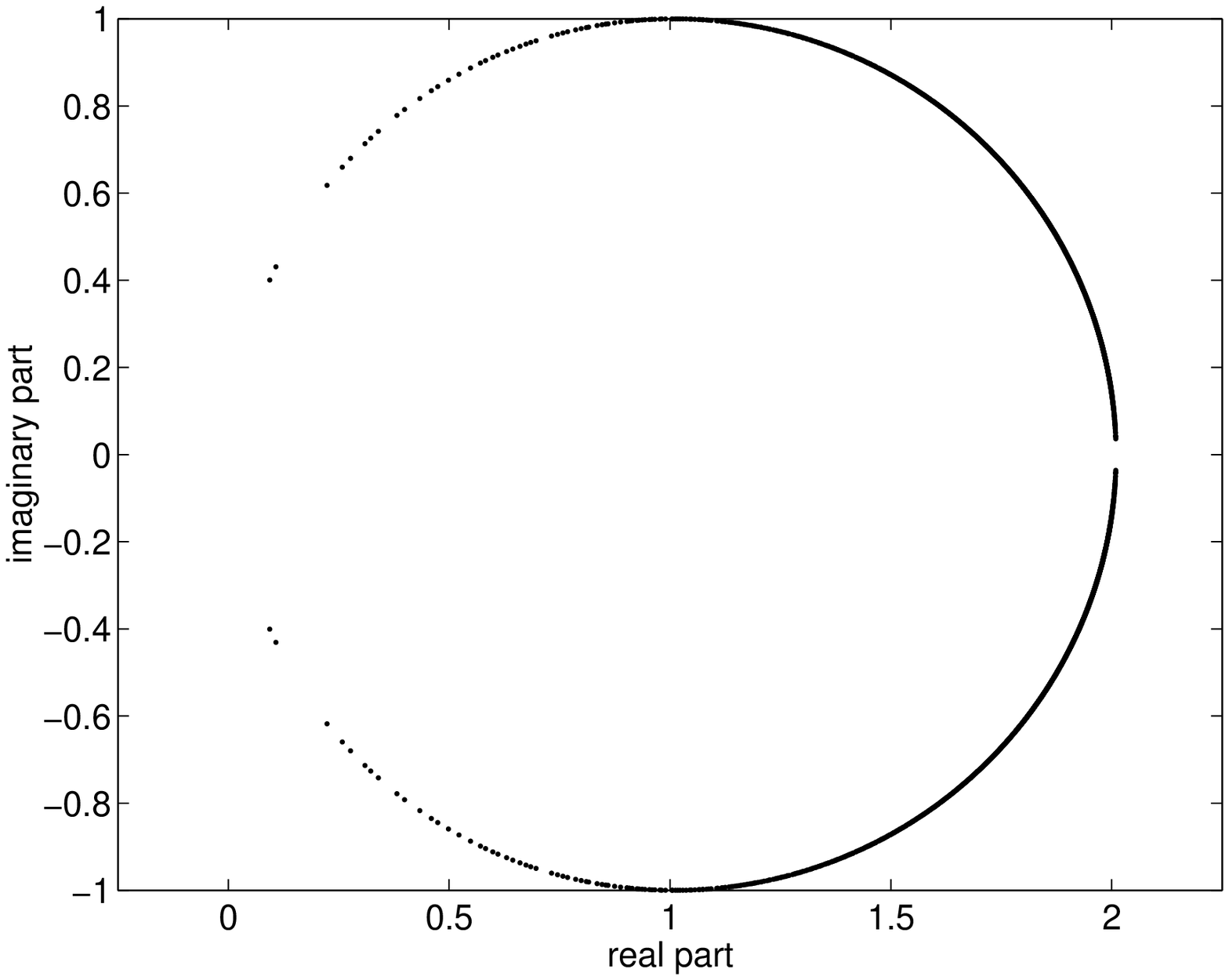}
\hfill
\includegraphics[scale=0.33]{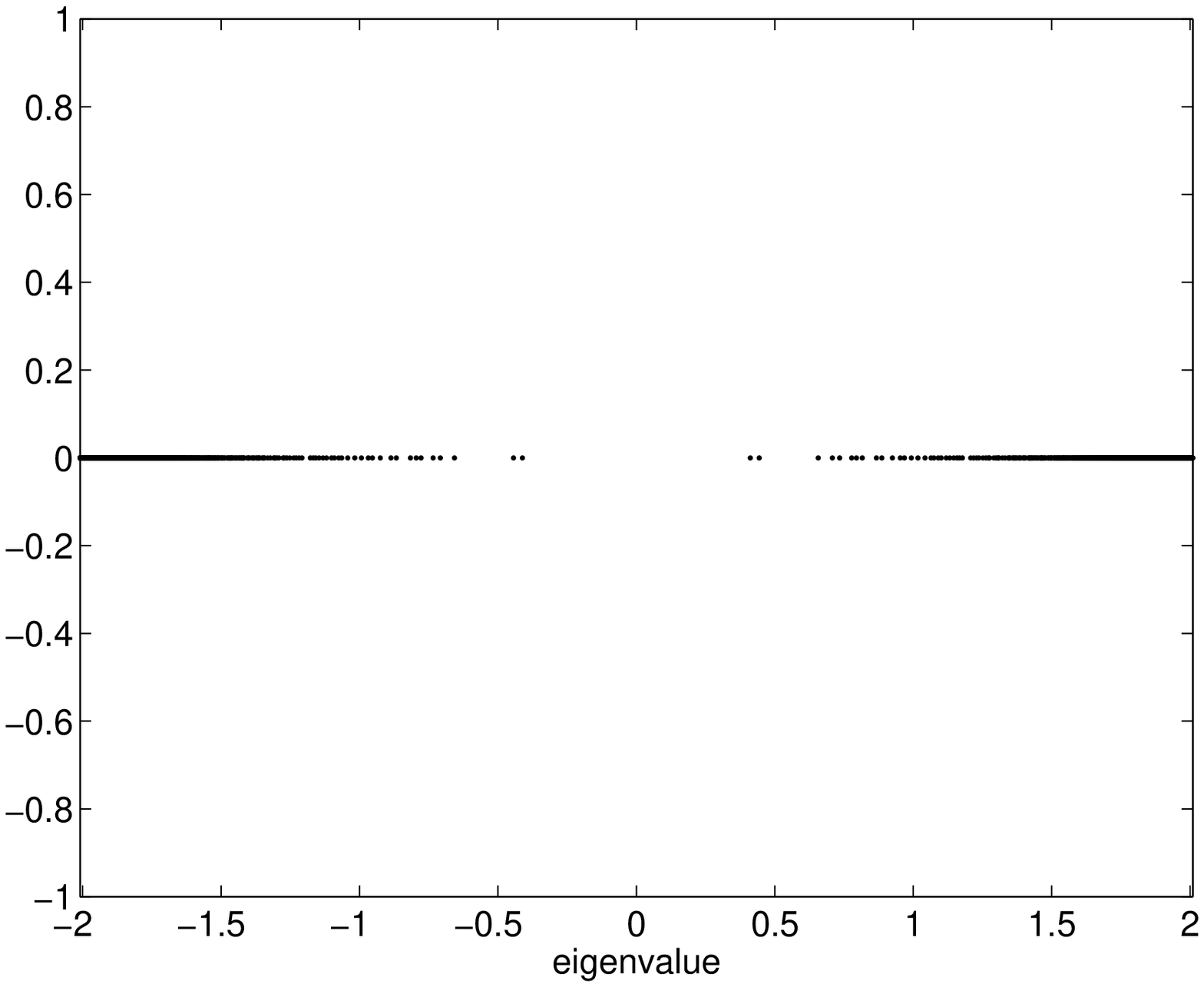}\\
\includegraphics[scale=0.33]{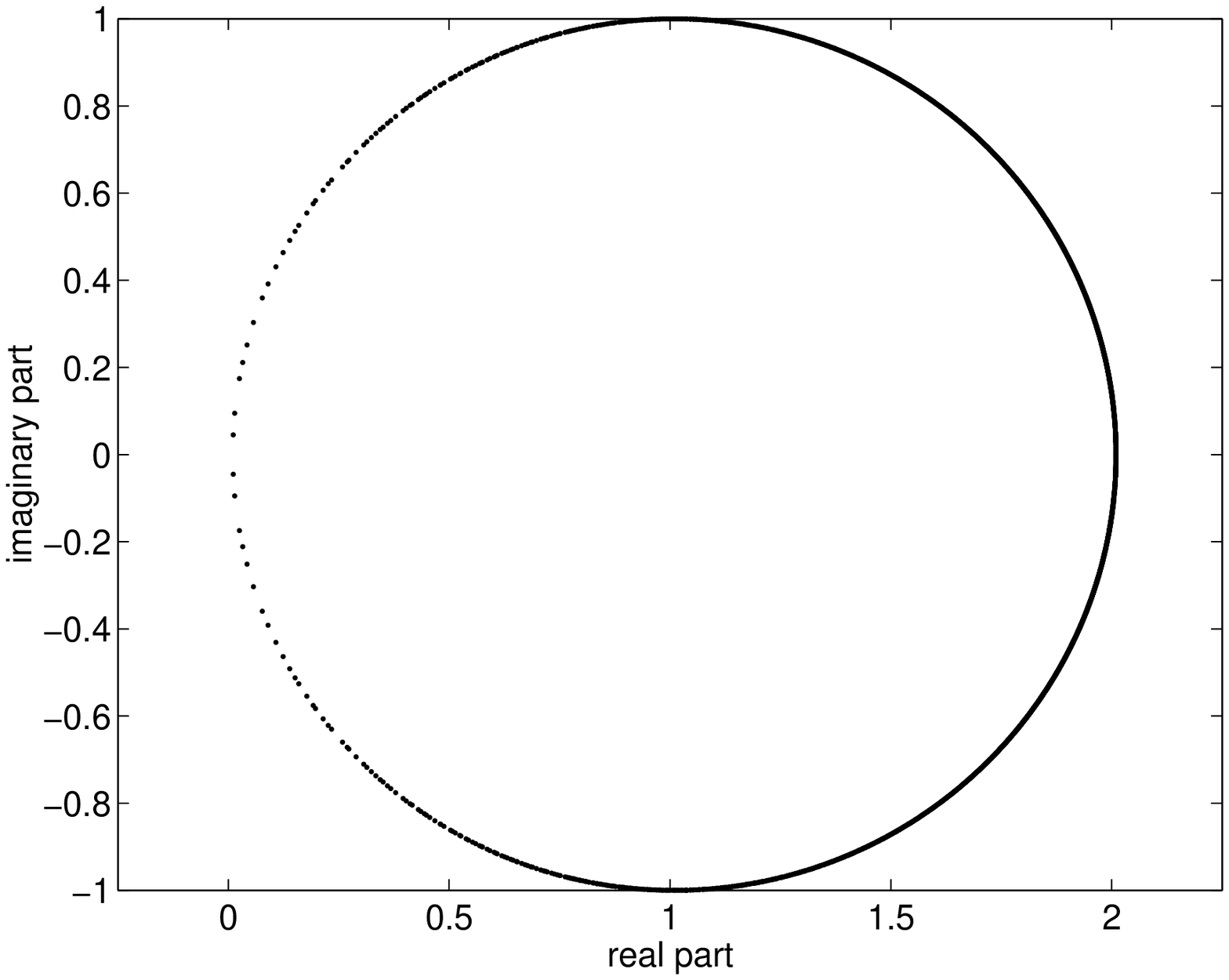}
\hfill
\includegraphics[scale=0.33]{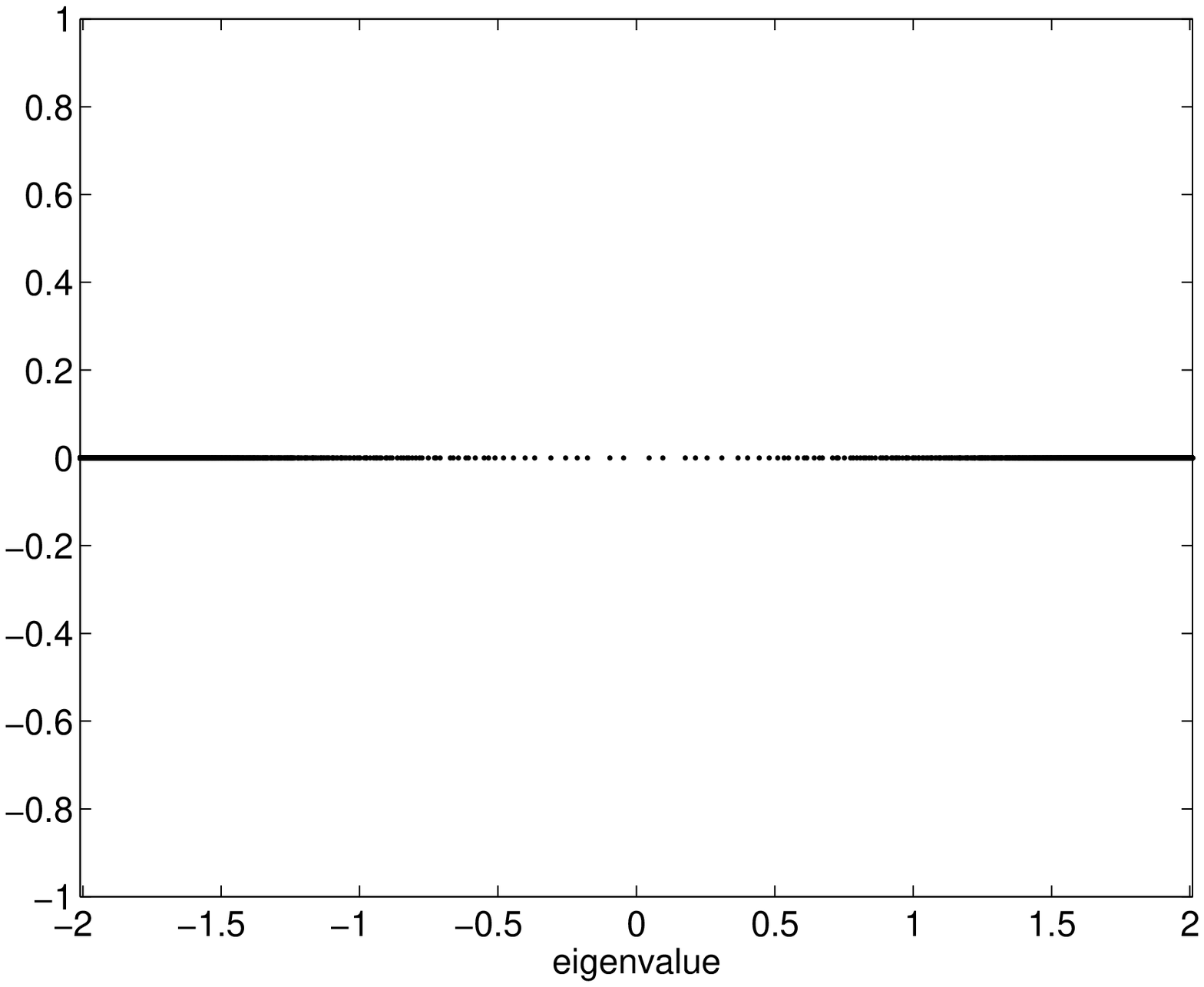}
\caption{Eigenvalues of $D_u$ (left column) and $D_h$ (right column) all for $\rho = 1.01$. The upper plots are for configuration
A, the lower for configuration B.}
\label{U_fig}
\end{figure}

Figure~\ref{U_fig} gives plots of the eigenvalues of $D_u$ and $D_h$ for our example
configurations with $\rho = 1.01$. It illustrates some interesting consequences
of Lemma~\ref{specdu_lem} and Lemma~\ref{specdh_lem}: With $C(\rho,1)$ denoting the
circle in the complex plane with radius 1 centered at $\rho$, we have
\begin{equation} \label{unitaryspec}
\spec(D_u) \subseteq C(\rho,1),
\end{equation}
and, moreover, $\spec(D_u)$ is symmetric w.r.t. the real axis.
On the other hand, $\spec(D_h)$ is ``almost symmetric'' w.r.t. the origin,
the only exceptions corresponding to the case $\sigma_j$ = 0, where
an eigenvalue of the form $\rho + \phi_j$
with $\phi_j \in \{ \pm1\}$ not necessarily matches $-\rho + \psi_j$ with $\psi_j
\in \{\pm1\}$, since we do not necessarily have $\psi_j = - \phi_j$.
Moreover,
\begin{equation} \label{hermitianspec}
\spec(D_h) \subseteq [-(\rho + 1), - (\rho -1)]
\cup [(\rho -1), (\rho +1)].
\end{equation}
Finally, let us note that we have some $\sigma_j$ equal to zero as soon
as 0 is an eigenvalue of the massless operator $\gamma_5 + \sign{Q}$, i.e.,\
as soon as the configuration has non-trivial topology.

\subsection{Dynamical simulations}
In a simulation with dynamical fermions, the costly computational task is
the inclusion of the fermionic part of the action into the ``force'' evolving
the gauge fields. This requires to solve the
``squared'' system
\[
D_u^{\dagger}D_u x = b \; \Longleftrightarrow \; D_h^2x = b.
\]
We will denote $D_n$ the respective operator, i.e.,\
\[
D_n = D_h^2 = D_u^{\dagger}D_u.
\]
If we plug in the definition of $D_u$, we find that
\begin{equation} \label{normal_eq}
  D_ny = D_u^{\dagger} D_u y =  \left( (\rho^2+1)I + \rho(\gamma_5\sign{Q}
 + \sign{Q}\gamma_5)\right) y = b,
\end{equation}
from which we get the interesting relation
\begin{equation} \label{2normal_eq}
  D_u D_u^\dagger = \gamma_5 D_u^\dagger D_u \gamma_5 = D_u^\dagger D_u.
 \end{equation}
As is well known, \eqnref{normal_eq} becomes block diagonal, since,
using (\ref{gamma5blockeq}) and (\ref{signblockeq}), we have
\begin{equation} \label{decoupled_eq}
D_n = 
\left( \begin{array}{cc} (\rho^2+1)I+2 \rho S_{11} & 0 \\
                                    0 & (\rho^2+1)I - 2 \rho S_{22}
                 \end{array}
            \right) y = b.
\end{equation}
 
In practical simulations, the decoupled structure of
(\ref{decoupled_eq})
can be exploited by running two seperate CG processes simultaneously, one for each (half-size)
block. Since each of these CG processes only has to accomodate a part of the spectrum of
$D_n$, convergence will be faster. An alternative usage of the block structure is to
compute the action of the matrix sign function on a two-dimensional subspace corrresponding to the two blocks and to use a block type Krylov subspace method. We do not, however, pursue this aspect
further, here.  As another observation, let us note that if $\gamma_5 b = \pm b$, then computing
$D_nb$ requires only one evaluation of the sign function instead of two (``chiral projection approach'').

As with the other formulations, let us summarize the important spectral properties of
the squared operator $D_n$.

\begin{lemma}
With the notation of Lemma~\ref{dec_lem}, $\spec(D_n) = \{
     \lambda_{j,\pm}^n\}$ with
\begin{eqnarray*}
  & \lambda^n_{j,\pm} = 1 + 2 \phi_j \rho + \rho^2 \enspace \mbox{(double
             eigenvalue) } &\quad  \mbox{ if }\quad
          \sigma_j \not = 0 \\
 & \lambda_{j,+} = (\rho + \phi_j)^2, \; \lambda_{j,-} =
       (-\rho + \psi_j)^2 &\quad \mbox{ if }\quad \sigma_j = 0.
\end{eqnarray*}
The corresponding eigenvectors are the same as for $D_h$ or, equivalently,
the same as for $D_u$ or, again equivalently,
\[
\left( \begin{array}{c}
                x^1_j \\
                 0
                \end{array}
                 \right), \;
             \left( \begin{array}{c}
                0 \\
                x^2_{j}
                \end{array}
                 \right) \; .
\]
\end{lemma}


Notice also that $\spec(D_n)$ satisfies
       \begin{equation} \label{normalspec}
           \spec(D_n) \subseteq [(\rho - 1)^2, (\rho +1)^2].
       \end{equation}

\section{Optimal Krylov subspace methods}\label{sec:met}
\subsection{Propagator computation}
Let us start with the non-hermitian matrix $D_u$.
Due to its shifted unitary form, there exists an
optimal Krylov subspace method based on short recurrences to solve \eqnref{unitary_eq}.
This method was published in \cite{JR94} and we would like to term it SUMR
(shifted unitary minimal residual). SUMR is mathematically equivalent to full GMRES, so its residuals
$r^m = b - D_ux^m$ at iteration $m$ are minimal in the 2-norm in the corresponding
affine Krylov subspace $x^0 +K_m(D_u,r^0)$ where  $K_m(D_u,r^0) = \span\{r^0,D_ur^0,\ldots,D_u^{m-1}r^0\}$.
From the algorithmic point of view,
SUMR is superior to full GMRES, since it relies on short recurrences and therefore
requires constant storage and an equal amount of arithmetic work
(one matrix vector multiplication and some vector operations) per iteration.
The basic idea of SUMR is the observation that the upper Hessenberg matrix
which describes the recursions of the Arnoldi process is shifted unitary so that 
its representation as a product of Givens rotations can be updated easily and with
short recurrences.
For the full algorithmic description we refer to \cite{JR94}.

Based on the spectral properties of $D_u$ that we identified in Section~\ref{sec:sys},
we can derive the following result on the convergence of
SUMR for \eqnref{unitary_eq}.
\begin{lemma} \label{sumr_lem}
Let $x^{k}$ be the $k$-th iterate of SUMR applied to
       \eqnref{unitary_eq} and let
       $r_u^{k}$ be its residual. Then the following estimate holds:
   \begin{equation} \label{sumr_est_eq}
      \| r_u^{k}\|_2 \leq 2 \cdot \left( \frac{1}{\rho} \right)^{k} \|r_u^0\|_2.
   \end{equation}
\end{lemma}
\begin{proof}
Since $D_u$ is normal, its field of values
$F(D_u) = \{ \langle D_ux, x \rangle, \|x \|_2=1 \}$ is the convex hull of 
its eigenvalues so
that we have $F(D_u) \subseteq C(\rho,1)$,
the disk centered at $\rho$ 
with radius 1. A standard result for full GMRES  (which is mathematically equivalent to SUMR) now
gives \eqnref{sumr_est_eq}, see, e.g., \cite{Gre97}.
\end{proof}

Let us proceed with the hermitian operator $D_h$ from \eqnref{hermitian_eq}, which is highly indefinite.
The MINRES method is the 
Krylov subspace method of choice for such systems:
It relies on short recurrences and it produces optimal iterates $x^k$
in the sense that their residuals $r^k_h = \gamma_5b - D_hx^k$ are minimal in the 2-norm
over the affine subspace $x^0 +K_m(D_h,r^0_h)$.

\begin{lemma} \label{minres_lem}
Let $x^{k}$ be the iterate of MINRES applied to \eqnref{hermitian_eq}
      at iteration $k$ and let
       $r_h^{k}$ be its residual. Then the following estimate holds:
   \begin{equation} \label{minres_est_eq}
      \| r_h^{k}\|_2 \leq 2 \cdot \left( \frac{1}{\rho} \right)^{\lfloor k/2 \rfloor} \|r_h^0\|_2.
   \end{equation}
   Here, $\lfloor k/2 \rfloor$ means $k/2$ rounded downwards to the nearest
   integer.
\end{lemma}
\begin{proof} \eqnref{minres_est_eq} is the standard
MINRES estimate (see \cite{Gre97}) with respect to the information from \eqnref{hermitianspec}.
\end{proof}

As a last approach to solving the propagator equation, let us consider the
standard normal equation
\begin{equation} \label{cgne_eq}
D_uD_u^\dagger z = b, \; x = D_u^\dagger z.
\end{equation}
Note that there exists an implementation of the CG method for \eqnref{cgne_eq}
known as CGNE \cite{GLo96}
which computes $x^k = D_u^\dagger z^k$ and its residual with respect to \eqnref{unitary_eq}
on the fly, i.e.,\ without additional work.


\subsection{Dynamical simulations}
We now turn to the squared equation \eqnref{normal_eq}. Since $D_n$ is hermitian and positive definite, the CG method
is the method of choice for the solution of \eqnref{normal_eq}, its iterates $y^m$ achieving minimal error in the energy norm (see Lemma~\ref{cg_lem} below) over the affine Krylov subspace
$y^0 + K_m(D_n,r^0)$.

\begin{lemma} \label{cg_lem}
Let $y^{k}$ be the iterate of CG applied to \eqnref{normal_eq}
         at stage $k$. Then the following estimates hold ($y^* = D_n^{-1}b$)
      \begin{eqnarray}
       \|y^k - y^*\|_{D_n} &\leq& 2 \cdot \left( \frac{1}{\rho} \right)^k \|y^0
               - y^*\|_{D_n},  \label{cg_energy_est_eq} \\
       \|y^k - y^*\|_2 &\leq& 2 \cdot \frac{\rho+1}{\rho-1} \left( \frac{1}{\rho} \right)^k \|y^0 
               - y^*\|_2. \label{cg_2norm_est_eq}
       \end{eqnarray}
       Here, $\| \cdot \|_{D_n}$ denotes the energy norm $\|y\|_{D_n} =
       \sqrt{y^{\dagger}D_ny}$.
\end{lemma}
\begin{proof}
The energy norm estimate \eqnref{cg_energy_est_eq}
is the standard estimate for the CG method based on the bound $\cond(D_n) \leq 
\left( (\rho+1)/(\rho-1) \right)^2$ for the condition number of $D_n$. The
$2$-norm estimate follows from the energy norm estimate using 
\[
\|y\|_2 \leq  \sqrt{\|D_n^{-1}\|_2} \cdot  \|y\|_{D_n} \leq \sqrt{\|D_n\|_2} \cdot \|y\|_2
\]
with
$\|D_n^{-1}\|_2 \leq 1/(\rho-1)^2, \; \|D_n\|_2 \leq (\rho+1)^2$.
\end{proof}

\section{Comparison of methods} \label{sec:comp}

Based on Lemma~\ref{sumr_lem} to \ref{cg_lem} we now proceed by theoretically investigating the
work for each of the three methods proposed so
far. We consider two  tasks: A propagator
computation where we compute the solution $x$ from \eqnref{unitary_eq} or \eqnref{hermitian_eq}, and a
dynamical simulation where we need to solve \eqnref{normal_eq}.

\subsection{Propagator computation}

The methods to be considered are
SUMR for \eqnref{unitary_eq}, MINRES  for \eqnref{hermitian_eq} and
CGNE for \eqnref{cgne_eq}. Note that due to \eqnref{2normal_eq}, Lemma~\ref{cg_lem}
can immediately also be applied to the CGNE iterates $z^k$ which approximate
the solution $z$ of \eqnref{cgne_eq}. In addition, expressing \eqnref{cg_energy_est_eq}
in terms of $x^k = D^\dagger_uz^k$ instead of $z^k$ turns
energy norms into 2-norms, i.e. we have
\begin{equation} \label{normal2est_eq}
    \|x^k - x^*\|_2 \leq 2 \cdot \left( \frac{1}{\rho} \right)^k \|x^0
               - x^*\|_2.
\end{equation}

In order to produce a reasonably fair account of how many
iterations we need, we fix a given accuracy $ \varepsilon$ for the final
error and calculate the
first iteration $ k (\varepsilon) $ for which the results given in Lemmas~\ref{sumr_lem} and
\ref{minres_lem} and in \eqnref{normal2est_eq} guarantee
\[
\parallel x^{k (\varepsilon)} - x^\ast \parallel_2 \, \leq
\varepsilon \cdot \parallel r^0 \parallel_2 , \quad x^\ast \mbox{ solution of \eqnref{unitary_eq}},
\]
where $ r^0 = b - D_u x^0$, $ x^0 $ being an identical starting vector for all
three methods (most likely, $ x^0 = 0 $). Since $ k (\varepsilon) $ will also
depend on $ \rho $, let us write $ k (\varepsilon, \rho) $. The following
may then be deduced from  Lemma~\ref{sumr_lem} and \ref{minres_lem} and
\eqnref{normal2est_eq} in a straightforward manner.

\begin{lemma}\label{comp_lem}
\begin{itemize}
\item[(i)] \ For SUMR we have
\[ \parallel x^k - x^\ast \parallel_2 \, \leq
\frac{1}{\rho - 1} \cdot \parallel r^k_u \parallel_2 \, \leq \frac{2}{\rho - 1}
\left( \frac{1}{\rho}\right)^k \cdot \parallel r^0 \parallel_2,
\]
 and therefore
\[
k (\varepsilon, \rho) \, \leq \frac{-\ln ( \varepsilon )}{\ln (\rho)} + \frac{-
\ln (2/(\rho - 1))}{\ln (\rho)} .
\]
\item[(ii)] \ For MINRES we have using $\|r_h^0\| = \|r_u^0\|$, since $r_h^0 = \gamma_5 r_u^0$
\[
\parallel x^k - x^\ast \parallel_2 \, \leq
\frac{1}{\rho - 1} \cdot \parallel r^k_h \parallel_2 \, \leq \frac{2}{\rho - 1}
\cdot \left( \frac{1}{\rho}\right)^{\lfloor \frac{k}{2}  \rfloor} \cdot
\parallel r^0_h \parallel_2,
\]
 and therefore
\[
k (\varepsilon, \rho) \, \leq 2 \cdot\left( \frac{-\ln (\varepsilon)}{\ln
(\rho)} + \frac{-\ln (2/(\rho - 1))}{\ln (\rho)} \right) .
\]

\item[(iii)] \ For CGNE we have
\[
\parallel x^k - x^\ast \parallel_2 \,
\leq 2 \cdot \left( \frac{1}{\rho} \right)^k \cdot \parallel x^0 - x^\ast
\parallel_2 \, \leq 2 \cdot \left( \frac{1}{\rho} \right)^k \cdot \frac{1}{\rho -
1} \cdot \parallel r^0 \parallel_2 ,
\]
 and therefore
\[
k (\varepsilon, \rho) \, \leq \frac{-\ln  (\varepsilon)}{\ln (\rho)} + \frac{ -
\ln (2/(\rho - 1))}{\ln (\rho)} .
\]
\end{itemize}
\end{lemma}

The arithmetic work in all these iterative methods is completely dominated
by the cost for evaluating the matrix vector product $\sign{Q}v$.
MINRES and SUMR require one such
evaluation per iteration, whereas CGNE requires two. Taking this into account,
Lemma~\ref{comp_lem} suggests that MINRES and CGNE should require about the 
same work to achieve a given accuracy $\varepsilon$, whereas SUMR should
need only half as much work, thus giving a preference for SUMR. Of course,
such conclusions have to be taken very carefully:
In a practical computation,
the progress of the iteration will depend on the {\em distribution} of
the eigenvalues, whereas the numbers $k(\varepsilon,\rho)$ of Lemma~\ref{dec_lem}
were obtained by just using {\em bounds} for the eigenvalues. 
For large values of $\rho$ the theoretical factor two  between SUMR and
MINRES/CGNE on the other hand, can be understood heuristically by the observation that SUMR can already reduce
the residual significantly by placing one root of its corresponding polynomial in the center of
the disc $C(\rho,1)$ whereas for the hermitian indefinite formulation two roots are necessary in the
two separate intervals. For smaller values of $\rho$ the differences are expected to be smaller except for
eigenvalues of $D_u$ close to $\rho+1$.

\begin{figure}
\centering
\includegraphics[scale=0.32]{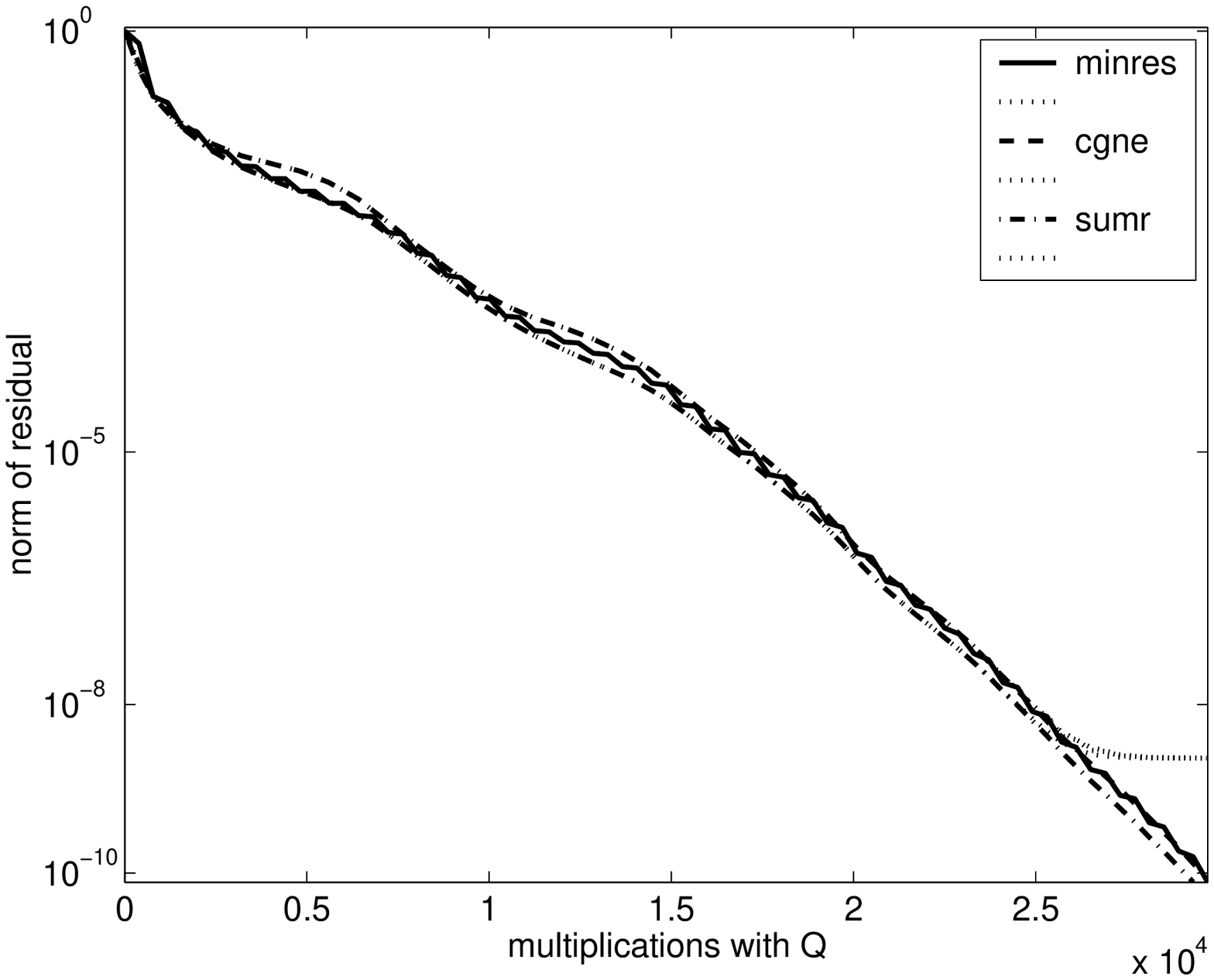}
\hfill \includegraphics[scale=0.32]{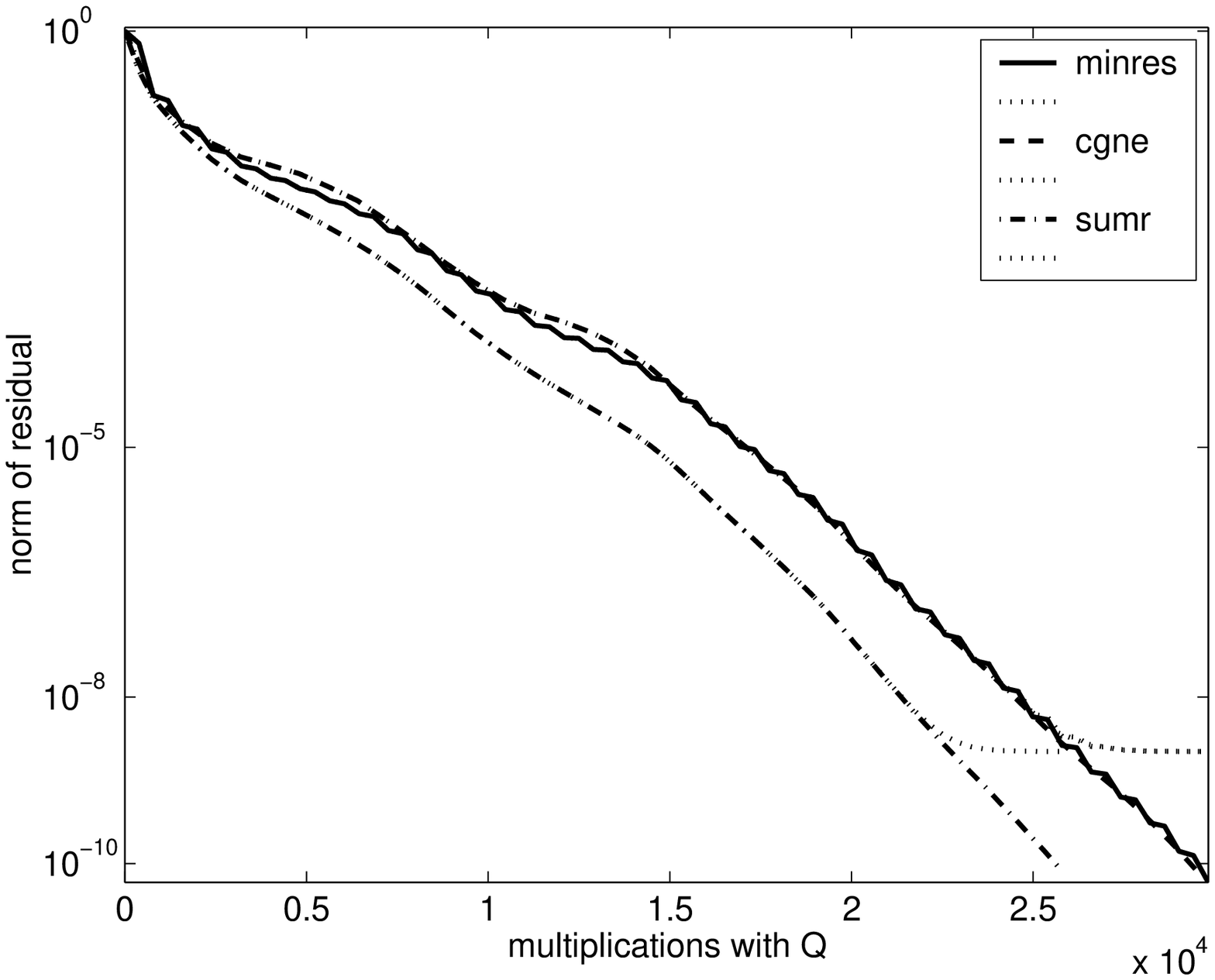}
\includegraphics[scale=0.32]{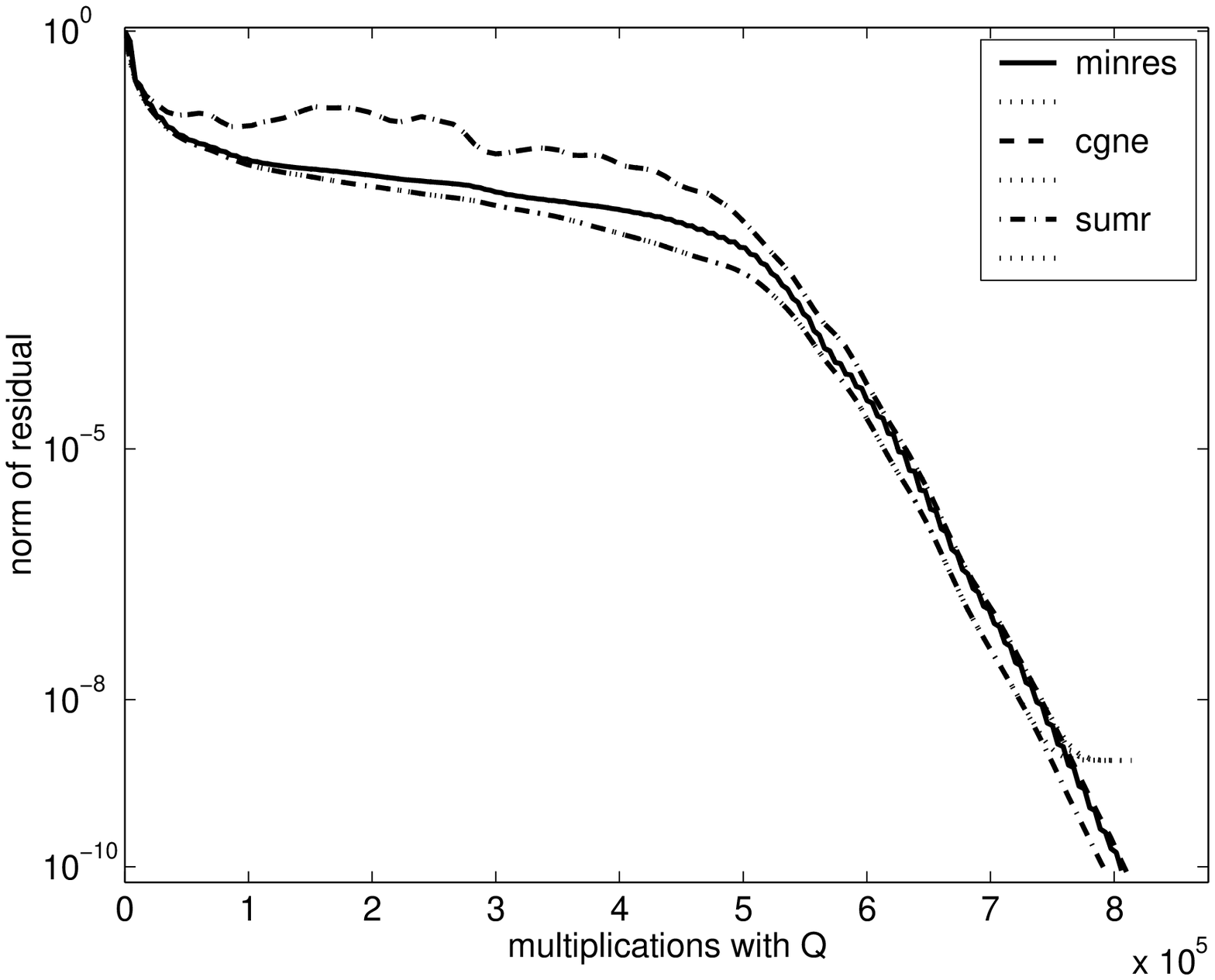}
\hfill \includegraphics[scale=0.32]{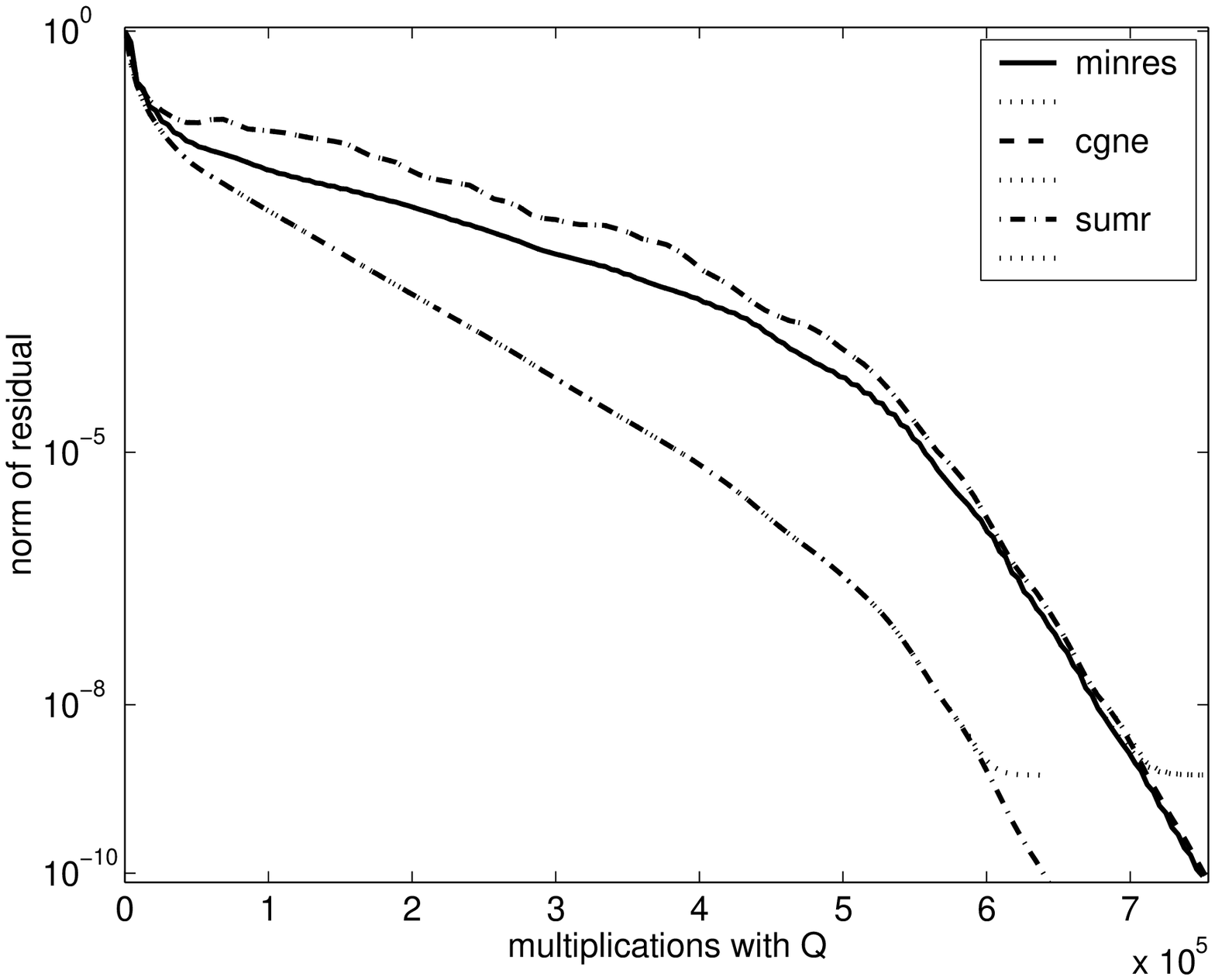}
\caption{Propagator computation: Convergence of MINRES for \eqnref{hermitian_eq},
CGNE for \eqnref{cgne_eq} and SUMR for
\eqnref{unitary_eq}. Left column is for $\rho =1.01$, right column for
$\rho = 1.1$. The upper plots are for configuration A, the lower for configuration B.}
\label{3methods_fig}
\end{figure}

Figure~\ref{3methods_fig} plots convergence diagrams for all three methods
for our example configurations. The diagrams
plot the relative norm of the residual $\|r^k\| / \|r^0\|$ as a function 
of the total number of matrix vector multiplications with the matrix $Q$.
These matrix vector multiplications represent the work for evaluating
$\sign{Q}v$ in each iterative step, since we use the multishift CG method
on a Zolotarev approximation with an accuracy of $10^{-8}$ with 10 poles (configuration A) and 20 poles (configuration B) respectively to
approximate $\sign{Q}v$. The true residual (dotted) converges to the accuracy of the inner iteration.
Note that the computations for configuration B are much more demanding, since the evaluation of $\sign Q\cdot v$ is more costly. We did not use any projection techniques to speed up this part of the computation.

\begin{figure}
\centering
\includegraphics[scale=0.32]{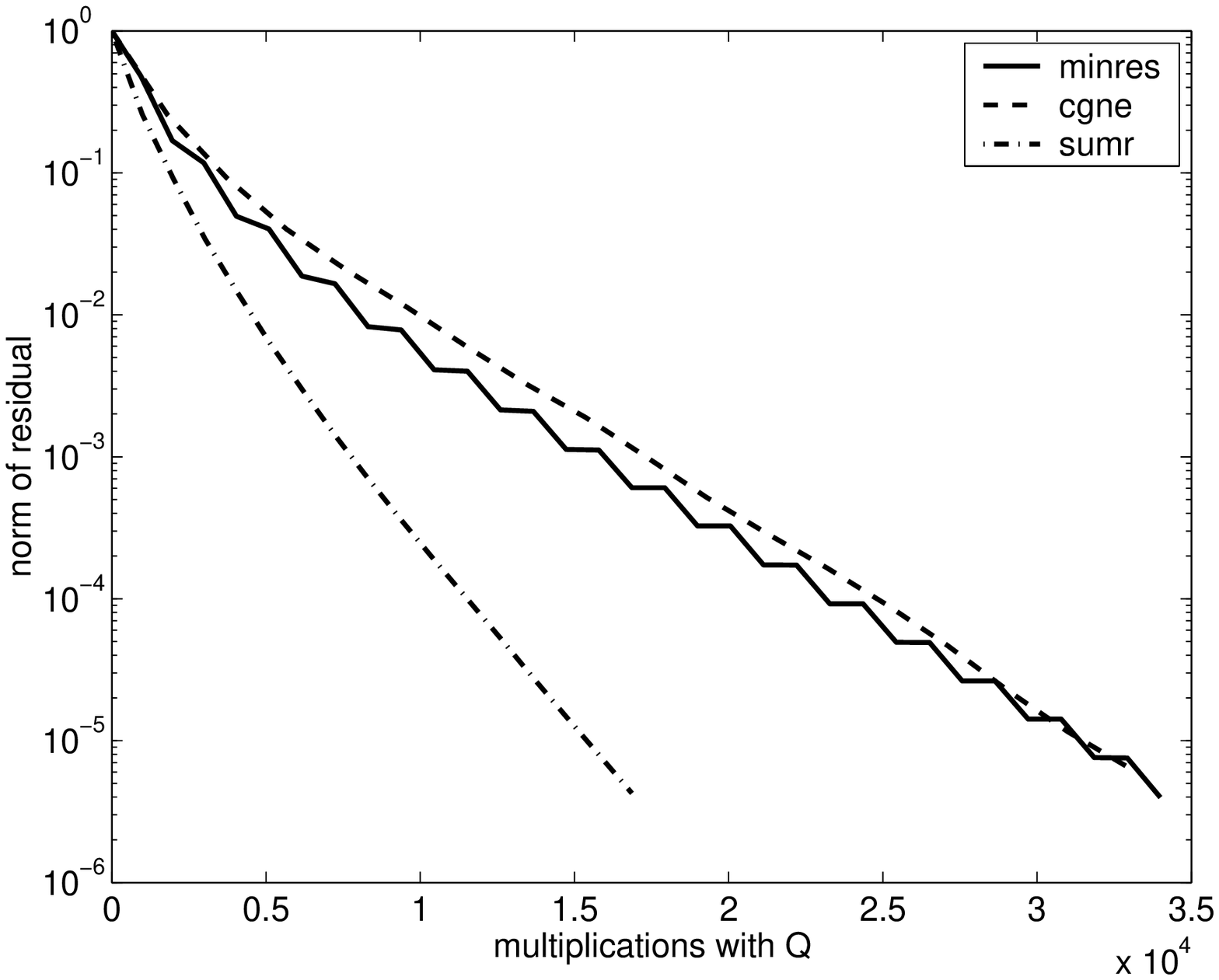}
\hfill \includegraphics[scale=0.32]{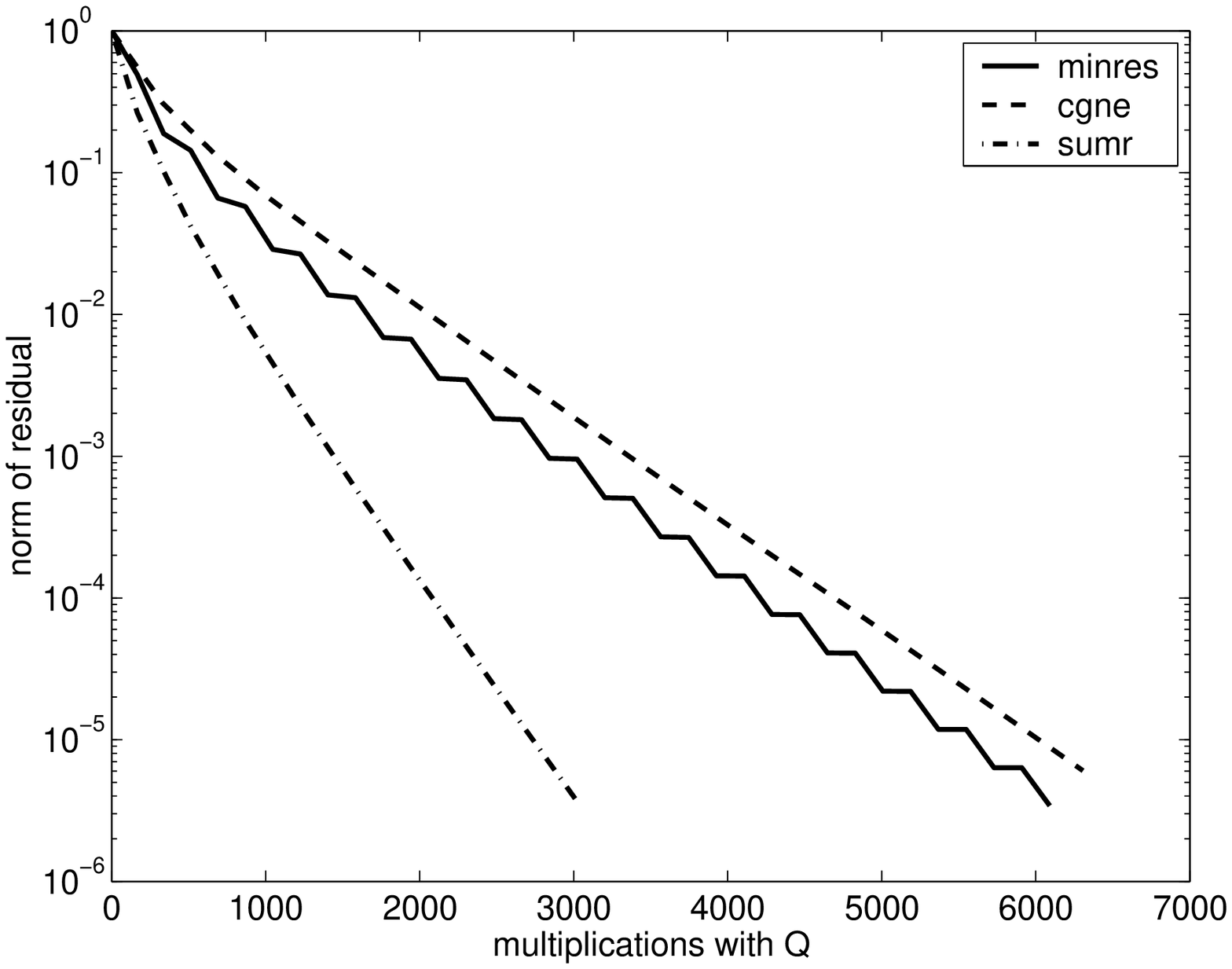}

\caption{Propagator computation: Convergence of MINRES for \eqnref{hermitian_eq},
CGNE for \eqnref{cgne_eq} and SUMR for
\eqnref{unitary_eq}. Left plot is for a $8^4$ lattice, right plot for a $16^4$ lattice.}
\label{3methods_fig2}
\end{figure}

Figure~\ref{3methods_fig2} plots convergence diagrams for all three methods for additional examples on
a $8^4$ lattice ($\beta = 5.6$, $\kappa = 0.2$, $\rho = 1.06$) and a $16^4$ lattice ($\beta = 6.0$, $\kappa = 0.2$, $\rho = 1.06$)  respectively.



MINRES and CGNE behave very similarly on configuration A. This is to be expected, since
the hermitian indefinite matrix $D_h$
is maximally indefinite, so that for an arbitrary right hand side
the squaring of the matrix inherent in $D_n$ should not significantly increase the
number of required iterations.

SUMR always performs best. The savings compared to MINRES and CGNE depend on $\rho$, $\beta$ and the lattice size and reached up to 50\%.

As a side remark, let us mention that the dichotomy between hermitian indefinite and
non-hermitian positive definite formulations is also a field of study in other areas.
For example, \cite{FRS98} investigates the effect of
multiplying some rows of a hermitian matrix with minus one in the context of
solving augmented systems of the form
\[
\left( \begin{array}{cc}
A & B\\
B^T & 0 \\
\end{array}
\right)
\left( \begin{array}{c}
x\\
y \\
\end{array}
\right)
=
\left( \begin{array}{cc}
b\\
 0 \\
\end{array}
\right)
\Longleftrightarrow
\left( \begin{array}{cc}
A & B\\
-B^T & 0 \\
\end{array}
\right)
\left( \begin{array}{c}
x\\
y \\
\end{array}
\right)
=
\left( \begin{array}{cc}
b\\
 0 \\
\end{array}
\right).
\]

\subsection{Dynamical simulations}

\begin{figure}
\centering
\includegraphics[scale=0.32]{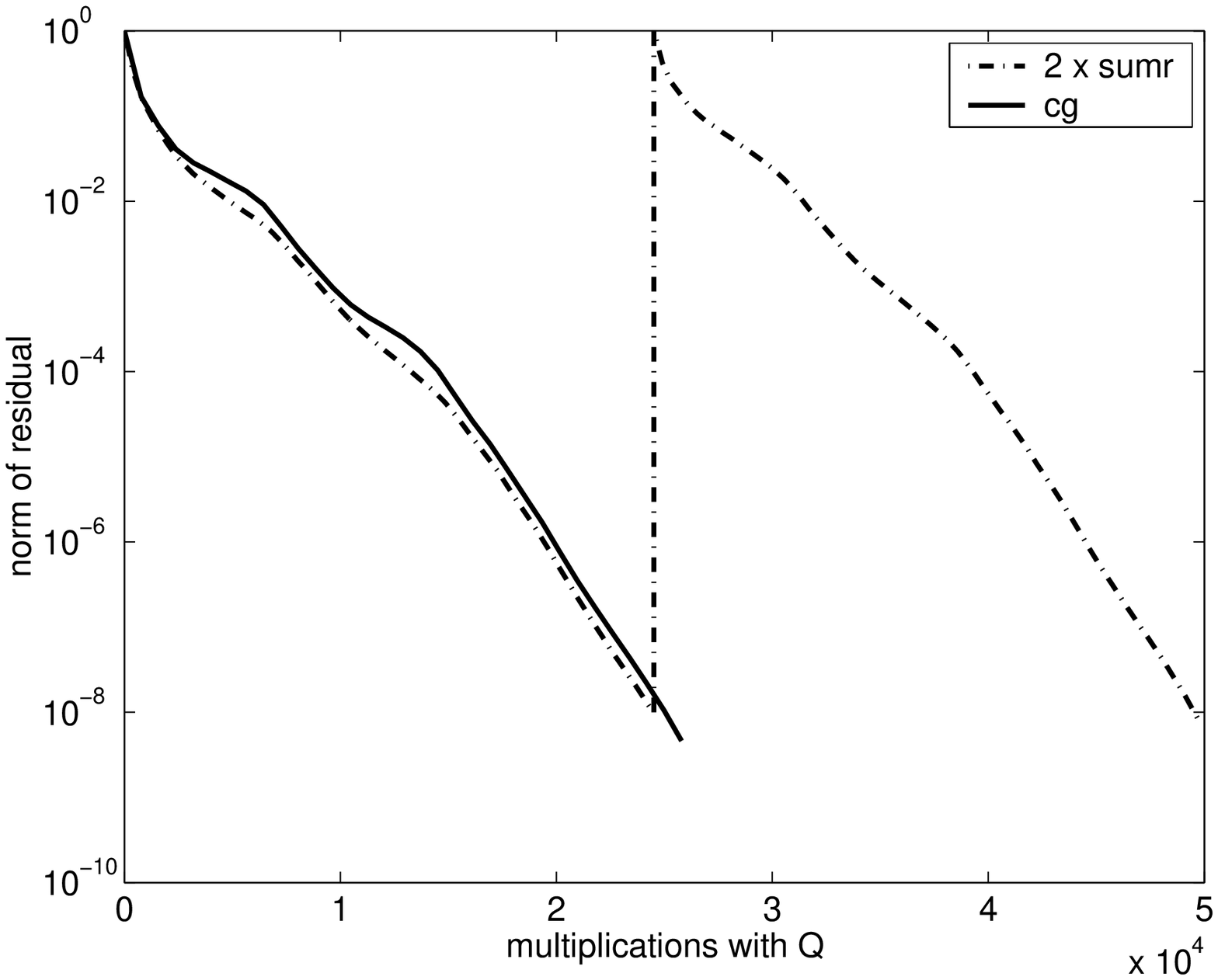}
\hfill \includegraphics[scale=0.32]{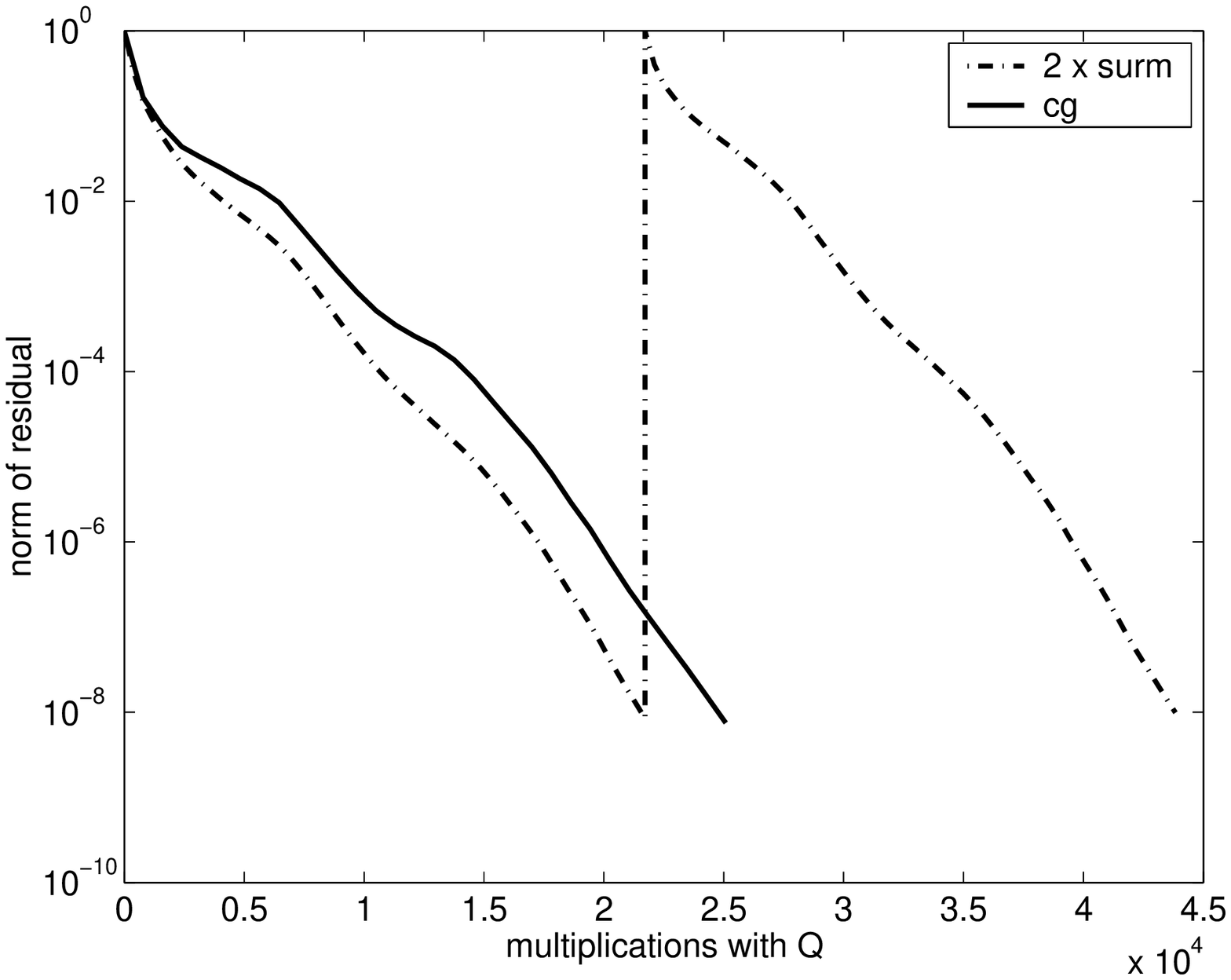}

\caption{Dynamical simulation: Convergence of two-sweep SUMR (dash-dotted) and CG (solid) for  \eqnref{normal_eq}
Left plot is for $\rho =1.01$, right plot for
$\rho = 1.1$. Both plots are for configuration A.}
\label{2methods_fig}
\end{figure}

Let us now turn to a dynamical simulation where we compute the solution $y^*$ from 
\eqnref{normal_eq}.
The methods to be considered are CG for \eqnref{normal_eq}, a two-sweep SUMR-approach
where we solve the two systems
\[
D_u^{\dagger} x = b, \quad D_uy = x
\]
using SUMR for both systems (note that $D_u^{\dagger}$ is of shifted unitary form, too), or
a two-sweep MINRES-approach solving the two systems 
\[
D_h x = b, \quad D_h y = x.
\]
It is now a bit more complicated to guarantee a comparable accuracy for each of the
methods. Roughly speaking, we have to run both sweeps in the two-sweep methods to
the given accuracy. Lemma~\ref{comp_lem} thus indicates that the two-sweep MINRES approach
will not be competitive, whereas it does not determine which of two-sweep SUMR or
CG is to be preferred. Our actual numerical experiments indicate that, in practice,
CG is superior to two-sweep SUMR, see Figure~\ref{2methods_fig}.

\section{Conclusion} \label{sec:con}
We have for the first time applied SUMR  as the outer iteration when solving a
propagator equation \eqnref{unitary_eq} for overlap fermions. Our theoretical
analysis and the numerical
results indicate that
this new method is superior to CGNE as well as to MINRES on
the symmetrized equation. In practice, savings tend to increase with larger values of $\rho$.
We achieved savings of about 50\% for physically interesting configurations.

We have also shown that when solving the squared equation \eqnref{normal_eq}
directly, as is required in a dynamical simulation, a two sweep approach using SUMR is
not competitive to directly using CG with the squared operator. This is in contrast
to the case of the Wilson fermion matrix, where it was shown in \cite{FHNLS94}
that
the two-sweep approach using BiCGstab performs better than CG.\footnote{BiCGstab is not
an alternative to SUMR in the case of the overlap operator, since, unlike BiCGstab,
SUMR is an optimal method achieving minimal residuals in the 2-norm.}

In the context of the overall inner-outer iteration scheme, additional issues arise.
In particular,
one should answer the question  of how accurately the result of the inner iteration (evaluating
the product of $\sign{Q}$ with a vector) is really needed. This issue will be addressed in
a forthcoming paper, \cite{ACE03}.

\medskip

{\bf Acknowledgements.} G.A. is supported under Li701/4-1 (RESH Forschergruppe FOR 240/4-1).
N.C. enjoys support from the EU Research and Training Network HPRN-CT-2000-00145 ``Hadron Properties from Lattice QCD''.

\appendix

\section{Definitions}

The Wilson-Dirac matrix reads $M = I - \kappa D_W$ where
\begin{equation}
(D_W)_{nm} =  \sum_{\mu} (I-\gamma_{\mu})\otimes
U_{\mu}(n)\delta_{n,m-\mu} + (I+\gamma_{\mu})\otimes
U^{\dagger}_{\mu}(n-\mu)\delta_{n,m+\mu}\label{eq:Diracmatrix}
\end{equation}
The Euclidean $\gamma$-matrices in the chiral representation are given
as:
\[
{\gamma_1} :=  \left[
{\begin{array}{rrrr}
0 & 0 & i &  0 \\
0 & 0 &  0 & i \\
-i & 0 & 0 & 0 \\
0 & -i& 0 & 0
\end{array}}
 \right]
{\gamma_2} :=  \left[
{\begin{array}{rrrr}
0 & 0 & -1 & 0 \\
0 & 0 & 0 & 1 \\
-1 & 0 & 0 & 0 \\
0  & 1 & 0 & 0
\end{array}}
 \right]
{\gamma_3} :=  \left[
{\begin{array}{rrrr}
0 & 0 &  0 & i \\
0 & 0 & -i & 0 \\
0 & i & 0 & 0 \\
-i &  0 & 0 & 0
\end{array}}
 \right]
{\gamma_4} :=  \left[
{\begin{array}{rrrr}
0 & 0 & 0 & 1 \\
0 & 0 & 1 & 0 \\
0 & 1 & 0 & 0 \\
1 & 0 & 0 & 0
\end{array}}
 \right] .
\]
The hermitian form of the Wilson-Dirac matrix is given by
\begin{equation}
Q=\gamma_5\, M,
\label{HWD}
\end{equation}
with $\gamma_5$ defined as the product
\begin{equation}
\label{GAMMA5}
\gamma_5:=
\gamma_1
\gamma_2
\gamma_3
\gamma_4
=\left[
{\begin{array}{rrrr}
1 & 0 & 0 & 0 \\
0 & 1 & 0 & 0 \\
0 & 0 & -1 & 0 \\
0 & 0 & 0 & -1
\end{array}}
 \right] .
\end{equation}

\section{Massive Overlap Operator}

Following Neuberger ~\cite{hep-lat/9710089}, one can write the massive
overlap operator as
\begin{equation}
  D_N(\mu) = c\left((1+\mu)I + (1-\mu)M (M^{\dagger}
    M)^{-\frac{1}{2}}\right)\label{eq:overlapoperator}.
\end{equation}
The normalisation $c$ can be absorbed into the fermion renormalisation, and
will not contribute to any physics. For convenience, we have set $c=1/(1-\mu)$.
Thus, the regularizing parameter $\rho$ as defined in \eq{overlap:def} is
related to $\mu$ by
\begin{equation}
\rho= (1+\mu)/(1-\mu).
\label{MASSDEFINITION}
\end{equation}

The physical mass of the fermion is then given by
\begin{equation}\label{FERMIONMASS}
m_f = \frac{2 \mu}{\kappa(1-\mu)}.
\end{equation}


\bibliographystyle{plain}
\bibliography{bibfile}

\begin{thebibliography}{10}

\bibitem{ACE03}
G.~Arnold, N.~Cundy, J.~{van den Eshof}, A.~Frommer, S.~Krieg, Th. Lippert, and
  K.~Sch\"afer.
\newblock Numerical methods for the {QCD} overlap operator: {III}. nested
  iterations.
\newblock to appear.

\bibitem{EHN98}
R.~G. Edwards, U.~M. Heller, and R.~Narayanan.
\newblock A study of practical implementations of the overlap-dirac operator in
  four dimensions.
\newblock {\em Nucl. Phys.}, B540:457--471, 1999.

\bibitem{FRS98}
B.~Fischer, A.~Ramage, D.~J. Silvester, and A.~J. Wathen.
\newblock Minimum residual methods for augmented systems.
\newblock {\em BIT}, 38(3):527--543, 1998.

\bibitem{FHNLS94}
A.~Frommer, V.~Hannemann, B.~N\"ockel, Th.\ Lippert, and K.~Schilling.
\newblock Accelerating wilson fermion matrix inversions by means of the
  stabilized biconjugate gradient algorithm.
\newblock {\em Int.\ J. of Mod.\ Phys.\ {\bf C}}, 5(6):1073--1088, 1994.

\bibitem{Glassner:1996gz}
U.~Gl{\"a}ssner, S.~G{\"u}sken, Th. Lippert, G.~Ritzenh{\"o}fer, K.~Schilling,
  and A.~Frommer.
\newblock How to compute {G}reen's functions for entire mass trajectories
  within {K}rylov solvers.
\newblock {\em Int. J. Mod. Phys.}, C7:635, 1996.

\bibitem{GLo96}
G.~H. Golub and C.~F. {Van Loan}.
\newblock {\em Matrix Computations}.
\newblock The John Hopkins University Press, Baltimore, London, 3rd edition,
  1996.

\bibitem{Gre97}
A.~Greenbaum.
\newblock {\em Iterative Methods for Solving Linear Systems}, volume~17 of {\em
  Frontiers in Applied Mathematics}.
\newblock Society for Industrial and Applied Mathematics (SIAM), Philadelphia,
  PA, 1997.

\bibitem{JR94}
C.~F. Jagels and L.~Reichel.
\newblock A fast minimal residual algorithm for shifted unitary matrices.
\newblock {\em Numer. Linear Algebra Appl.}, 1(6):555--570, 1994.

\bibitem{Narayanan:2000qx}
R.~Narayanan and H.~Neuberger.
\newblock An alternative to domain wall fermions.
\newblock {\em Phys. Rev.}, D62:074504, 2000.

\bibitem{hep-lat/9710089}
H.~Neuberger.
\newblock Vector like gauge theories with almost massless fermions on the
  lattice.
\newblock {\em Phys. Rev.}, D57:5417--5433, 1998.

\bibitem{PWe94}
C.~C. Paige and M.~Wei.
\newblock History and generality of the {${\rm CS}$} decomposition.
\newblock {\em Linear Algebra Appl.}, 208/209:303--326, 1994.

\bibitem{EFL02}
J.~{van den Eshof}, A.~Frommer, Th. Lippert, K.~Schilling, and H.A. {van der
  Vorst}.
\newblock Numerical methods for the {QCD} overlap operator: {I}. sign-function
  and error bounds.
\newblock {\em Comput. Phys. Comm.}, 146:203--224, 2002.

\end{thebibliography}



\printindex
\end{document}